\documentclass[submission,copyright,creativecommons]{eptcs}
\providecommand{\event}{2011 Refinement Workshop} 
\usepackage{breakurl}             
\usepackage{subfigure}
\usepackage{color}
\usepackage{multirow}
\usepackage{bsymb,b2latex}
\usepackage{graphicx}
\usepackage{arydshln}
\usepackage{mdwlist}

\title{Discovery of Invariants through \\Automated Theory Formation \footnote{
The research reported in this paper is supported by EPSRC grants
EP/F037058 and EP/F035594.}}

\author{Maria Teresa Llano\thanks{
Maria Teresa Llano is partially funded by a BAE Systems studentship.} \qquad Andrew Ireland
\institute{Heriot-Watt University\\}
\institute{School of Mathematical and Computer Sciences\\}
\and
Alison Pease 
\institute{University of Edinburgh\\}
\institute{School of Informatics\\}
}
\def\titlerunning{Discovery of Invariants through Automated Theory Formation}
\def\authorrunning{M.T. Llano, A. Ireland \& A. Pease}

\begin{document}
\maketitle

\begin{abstract}
Refinement is a powerful mechanism for mastering the complexities 
that arise when formally modelling systems. Refinement also brings
with it additional proof obligations -- requiring a developer to 
discover properties relating to their design decisions. 
With the goal of reducing this burden, we have investigated how
a general purpose theory formation tool, HR, can be used to 
automate the discovery of such properties within the context of
Event-B. Here we develop a heuristic 
approach to the automatic discovery of invariants and report upon a 
series of experiments that we undertook in order to evaluate our approach.
The set of heuristics developed provides 
systematic guidance in tailoring HR for a given Event-B development. 
These heuristics are based upon proof-failure analysis, and have
given rise to some promising results. 
\end{abstract}

\section{Introduction}
\label{sec:introduction}

By allowing a developer to incrementally introduce design details, refinement 
provides a powerful mechanism for mastering the complexities that arise when
formally modelling systems. This benefit comes with proof obligations (POs) -- the task 
of proving the correctness of each refinement step. Discharging such proof obligations
typically requires a developer to supply properties -- properties that relate
to their design decisions. Ideally automation should be provided to support 
the discovery of such properties, allowing the developer to focus on design 
decisions rather than analysing failed proof obligations.

\begin{figure}[h]
\begin{center}
\includegraphics[scale=0.5]{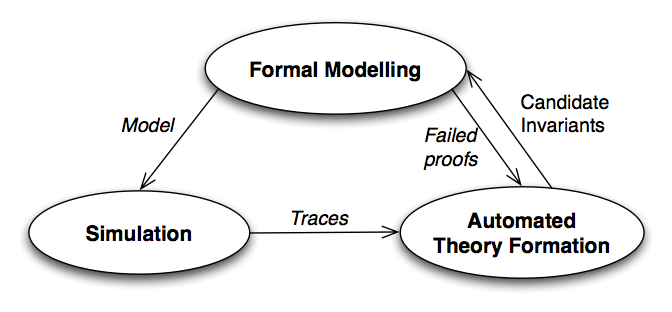}
\end{center}
\caption{Approach for the automatic discovery of invariants.}
\label{Approach}
\end{figure}

With this goal in mind, we have developed a heuristic approach for the automatic discovery of
invariants in order to support the formal modelling of systems.  Our approach, shown
in Figure \ref{Approach}, involves three components:

\begin{itemize}
 \item a simulation component that generates system traces,
 \item an Automatic Theory Formation (ATF) component that generates conjectures from the analysis of the traces and,
 \item a formal modelling component that supports proof and proof failure analysis.
\end{itemize}

Crucially, proof and proof failure analysis is used to tailor the theory formation component.

From a modelling perspective we have focused on Event-B \cite{EventbBook} and the Rodin 
tool-set \cite{Rodin}, in particular we have used the ProB animator plug-in \cite{ProB03} for the
simulation component.  In terms of ATF, we have used a general-purpose system called HR
\cite{colton:book}.  Generating invariants from the analysis of ProB animation traces is an approach
analogous to that of the Daikon system \cite{ErnstPGMPTX2007}; however, while Daikon is
tailored for programming languages here we focus on formal models.  We come back to this in \S\ref{sec:future}.
 
Our investigation involved a series of experiments, drawing upon 
examples which include Abrial's ``Cars on a Bridge'' \cite{EventbBook} and the Mondex case 
study by Butler et al. \cite{Mondex} . Our initial experiments highlighted the power of
HR as a tool for automating the discovery of both system and gluing invariants -- system invariants 
introduce requirements of the system while gluing invariants 
relate the state of the refined model with the state of the abstract model.  However,  
our experiments also showed significant limitations: i) selecting the right configuration
for HR according to the domain at hand, i.e. selection of production rules and 
the number of theory formation steps needed to generate the missing invariants, and ii)
the overwhelming number of conjectures that are generated. This led us to consider how HR could be 
systematically tailored to provide practical support during an Event-B
development. As a result we developed a set of heuristics which are based
upon proof-failure analysis. These heuristics have given rise to some
promising results and are the main focus of this paper.  Although we 
show here the application of our technique in the context of Event-B, we believe our approach can be 
applied to any refinement style formal modelling framework that
supports simulation and that uses proof in order to verify refinement steps. 

The remainder of this paper is organised as follows. 
In \S\ref{sec:background} we provide background on both
Event-B and HR. The application of HR within the context of
Event-B is described in \S\ref{sec:aft4eventB}, along with the
limitations highlighted above. In \S\ref{sec:approach}
we present our heuristics, and describe their rationale. 
Our experimental results are given in \S\ref{sec:experiments}, 
while related and future work are discussed in \S\ref{sec:future}.

\section{Background}
\label{sec:background}

\subsection{Event-B} \label{EventB-Background}
Event-B promotes an incremental style
of formal modelling, where each step of a development is underpinned by formal 
reasoning.  An Event-B development is structured around {\em models} and {\em contexts}.
A context represents the static parts of a system, i.e. {\em constants} and {\em axioms},
while the dynamic parts are represented by models. Models have a state, i.e. 
{\em variables}, which are updated via guarded actions, known as {\em events}, and 
are constrained by {\em invariants}. 

\begin{figure}[h]
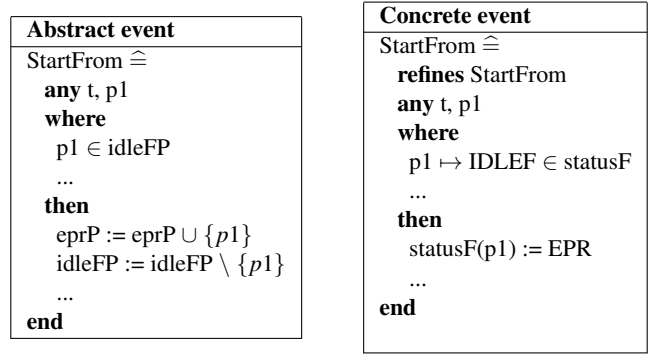

\begin{center}
\begin{footnotesize}
\begin{tabular}{|l|}
\hline
\textbf{Abstract event} \\
\hline
StartFrom $\widehat{=}$ \\
~~ \textbf{any} t, p1 \\
~~ \textbf{where} \\
~~~~ p1 $\in$ idleFP \\
~~~~ ... \\
~~ \textbf{then} \\
~~~~ eprP := eprP $\cup$ $\{p1\}$ \\
~~~~ idleFP := idleFP $\setminus$ $\{p1\}$ \\
~~~~ ... \\
\textbf{end} \\
\hline
\end{tabular}  \ \hspace*{0.2in} \
\begin{tabular}{|l|}
\hline
\textbf{Concrete event} \\
\hline
StartFrom $\widehat{=}$ \\
~~ \textbf{refines} StartFrom \\
~~ \textbf{any} t, p1 \\
~~ \textbf{where} \\
~~~~ p1 $\mapsto$ IDLEF $\in$ statusF \\
~~~~ ... \\
~~ \textbf{then} \\
~~~~ statusF(p1) := EPR \\
~~~~ ... \\
\textbf{end} \\
\\
\hline
\end{tabular}
\end{footnotesize}
\end{center}
\caption{Abstract and concrete views of event \emph{StartFrom}.}
\label{Events}
\end{figure}

To illustrate the basic features of a refinement consider the two events shown in 
Figure \ref{Events}, which are part of the Mondex development \cite{Mondex}.  The Mondex 
system models the transfer of money between electronic purses.  The event {\em StartFrom} 
handles the initiation of a transaction on the side of the source purse. In order to 
initiate a transaction, the source purse must be in the {\em idle} state (waiting state) and after
the transaction has been initiated the state of the purse must be changed to {\em epr} (expecting request).  
As shown in Figure \ref{Events}, in this step of the refinement the abstract model represents the state of purses by disjoint 
sets, i.e. the variables {\em eprP} and {\em idleFP}, while the concrete model handles these states through a function, 
i.e. the variable {\em statusF}, which maps a purse to an enumerated set that represents the current state, i.e. the 
constants {\em IDLEF} and {\em EPR}.

Note that the keyword {\bf refines} specifies the event being refined, while the keywords {\bf any}, {\bf where} and {\bf then} delimit event 
{\em parameters}, {\em guards} and {\em actions} respectively. Note also that the concrete event on the right represents a refinement of the abstract event on the left.
 
In order to verify this refinement an invariant is 
required that relates the concrete and abstract states --
these are known as {\em gluing invariants}. In the case of the events 
given above, the required gluing invariant takes the form:
\begin{eqnarray}
\mbox{idleFP} = \mbox{statusF}^{-1}[\{\mbox{IDLEF\}}]
\label{GluingInvariant}
\end{eqnarray} 

This invariant states that the abstract set {\em idleFP} can be obtained from the inverse 
of the function {\em statusF} evaluated over the enumerated set {\em IDLEF}.
A similar gluing invariant would be required for the abstract set {\em eprP} and the function {\em statusF}. 
Within the Rodin toolset\footnote{Rodin provides an Eclipse based platform for Event-B, with a range of
modelling and reasoning plug-ins, {\em e.g.} UML-B \cite{UML-B06}, 
ProB model checker and animator \cite{ProB03}, B4free theorem prover (\url{http://www.b4free.com}).},
the user is required to supply such gluing invariants. Likewise,
invariants relating to state variables within a single model must also be
supplied by the user -- what we refer to here as {\em system invariants}. To 
illustrate, the following disjointness property represents an invariant
of the abstract event above:

\begin{eqnarray*}
\mbox{eprP} ~ \cap ~ \mbox{idleFP} = \o
\nonumber
\end{eqnarray*}

From a theoretical perspective such invariants are typically not very challenging. 
They are however numerous and represent a significant
obstacle to increasing the accessibility of formal refinement approaches 
such as Event-B. 

\subsection{Automated theory formation and HR} \label{HRBackground}
Lenat developed one of the earliest examples of a discovery system in mathematics; Automated Mathematician
(AM) \cite{lenat76} and its successor Eurisko \cite{lenat83}. Despite subsequent methodological
criticism of Lenat's work \cite{partridge}, he did show us that it is
possible to formalise heuristics for discovery in mathematics. Colton
has developed this intuition in his HR machine learning system\footnote{HR is named after
mathematicians Godfrey Harold Hardy (1877 - 1947) and Srinivasa
Aiyangar Ramanujan (1887 - 1920).} \cite{colton:book}. HR performs
descriptive induction to form a theory about a set of objects of interest which are described by a set of core
concepts (this is in contrast to predictive learning systems which are used to solve the
particular problem of finding a definition for a target concept). 
Based on Colton's observation that it is possible to gain an understanding of a complex concept
by decomposing it via small steps into simpler concepts, Colton 
defined production rules which take in concepts and make 
small changes to produce further concepts.

HR constructs a theory by finding examples of objects of interest, inventing new concepts, making plausible statements relating
those concepts, evaluating both concepts and statements and, if working in a mathematical 
domain, proving or disproving the statements.  Objects of interest are the entities which a theory discusses. For 
instance, in number theory the objects of interest are integers, in group theory they are groups, 
etc.  Concepts are either provided by the user (core concepts) or developed by HR (non-core concepts) and 
have an associated data table (or table of examples). The data table is a function
from an object of interest, such as the number 1, or the prime 3, to a 
truth value or a set of objects. 

Each production rule is generic and works by performing operations on the content of one or two input data tables and a set of parameterisations 
in order to produce a new output data table, thus forming a new concept. The production rules and parameterisations are usually 
applied automatically according to a search strategy which has been entered by the user, and are applied 
repeatedly until HR has either exhausted the search space or has reached a user-defined number of theory formation steps 
to perform.  Production rules include:

\begin{itemize}
\item The \it split \rm rule: this extracts the list of examples 
of a concept for which some given parameters hold.
\item The \it negate \rm rule: this negates predicates in the new
definition.
\item The \it compose \rm rule: combines predicates from two old 
concepts in the new concept.
\item The \it arithmetic \rm rule: performs arithmetic operations (+, -, $\ast$, $\div$)
on specified entries of two concepts.
\item The \it numrelation \rm rule: performs arithmetic comparisons ($<$, $>$, $\leq$, $\geq$)
on specified entries of two concepts.
\end{itemize}

Each time a new concept is generated, HR checks to see whether it can  
make conjectures with it. This could be equivalence conjectures, if the new concept 
has the same data table as a previous concept; implication conjectures, if the data table of 
the new concept either subsumes or is subsumed by that of another concept, or non-existence 
conjectures, if the data table for the new concept is empty.

Thus, the theories HR produces contain concepts which relate the objects of
interest; conjectures which relate the concepts; and proofs which
explain the conjectures. Theories are constructed via theory formation
steps which attempt to construct a new concept and, if successful, 
formulate conjectures and evaluate the results.  
HR has been used for a variety of discovery projects, including
mathematics and scientific domains (it has been particularly
successful in number theory \cite{colton:00} and algebraic domains
\cite{meier:calculemus02}) and constraint solvers \cite{colton:cp01,lsr09}.

\begin{figure}[h]
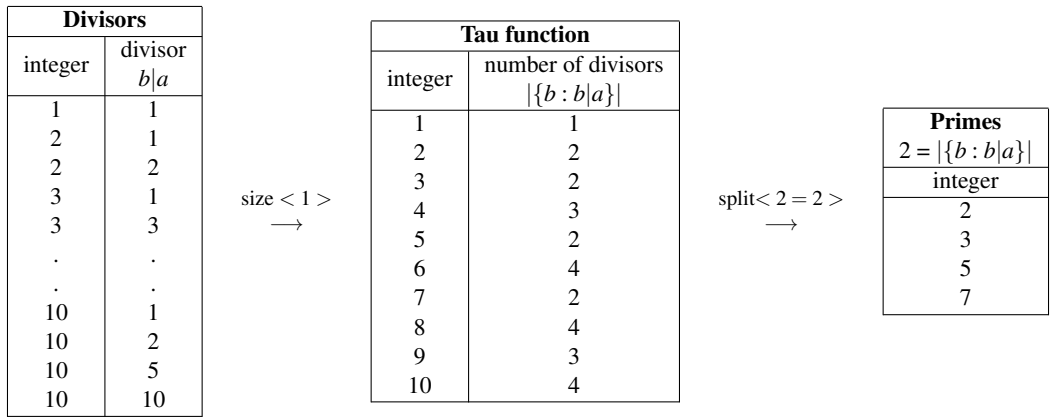

\begin{center}
\begin{footnotesize}
\subfigure{
\begin{tabular}{|c|c|}
\hline
\multicolumn{2}{|c|}{\textbf{Divisors}} \\
\hline
\multirow{2}{*}{integer} & divisor \\
& $b|a$ \\
\hline
1 & 1 \\
2 & 1 \\
2 & 2 \\
3 & 1 \\
3 & 3 \\
. & . \\
. & . \\
10 & 1 \\
10 & 2 \\
10 & 5 \\
10 & 10 \\
\hline
\end{tabular}
}
\subfigure{
\begin{scriptsize}
\begin{tabular}{c}
size $<1>$ \\
$\longrightarrow$ \\
\end{tabular}
\end{scriptsize}
}
\subfigure{
\begin{tabular}{|c|c|}
\hline
\multicolumn{2}{|c|}{\textbf{Tau function}} \\
\hline
\multirow{2}{*}{integer} & number of divisors \\
& $|\{b : b|a\}|$ \\
\hline
1 & 1 \\
2 & 2 \\
3 & 2 \\
4 & 3 \\
5 & 2 \\
6 & 4 \\
7 & 2 \\
8 & 4 \\
9 & 3 \\
10 & 4 \\
\hline
\end{tabular}
}
\subfigure{
\begin{scriptsize}
\begin{tabular}{c}
split$<2 = 2>$ \\
$\longrightarrow$ \\
\end{tabular}
\end{scriptsize}
}
\subfigure{
\begin{tabular}{|c|}
\hline
\textbf{Primes} \\
2 = $|\{b : b|a\}|$ \\
\hline
integer \\
\hline
2 \\
3 \\
5 \\
7 \\
\hline
\end{tabular}
}
\end{footnotesize}
\end{center}
\caption{Steps applied by HR to produce the concept of prime numbers.}
\label{prime}
\end{figure}

As an example, we show how HR produces the concept of prime numbers and the conjecture that all
prime numbers are non-squares.  Figure \ref{prime} shows the data tables used by HR for the formation of
the concept of prime numbers.  

In order to generate this concept, HR would take in the concept of divisors ($b|a$ 
where $b$ is a divisor of $a$), represented by a data table for a subset of integers 
(partially shown in Figure \ref{prime} for integers from 1 to 10). Then, HR would apply the size 
production rule with the parameterisation $<1>$.  This means that the number of tuples 
for each entry in column 1 are counted, and this number is then recorded for each entry. For 
instance, in the data table representing the concept of divisors, 1 appears only once in the first column, 
2 and 3 appear twice each, and 10 appears four times. This number is recorded next to the 
entries in a new data table (the table for the concept Tau function). HR then takes in this 
new concept and applies the split production rule with the parameterisation $<2 = 2>$, 
which means that it produces a new data table consisting of those entries in the previous data 
table whose value in the second column is 2. This is the concept of a prime number.

\begin{figure}[h]
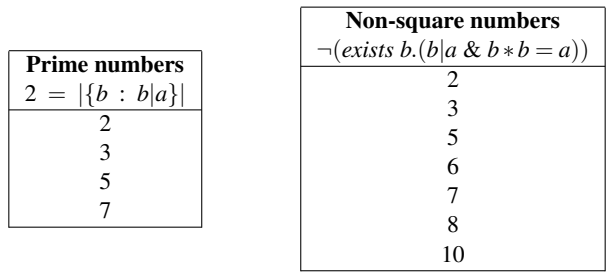

\begin{center}
\begin{footnotesize}
\subfigure{
\begin{tabular}{|c|}
\hline
\textbf{Prime numbers} \\
$2 ~ = ~ |\{b ~ : ~ b|a\}|$ \\
\hline
2 \\
3 \\
5 \\
7 \\
\hline
\end{tabular}
}
\hspace{1cm}
\subfigure{
\begin{tabular}{|c|}
\hline
\textbf{Non-square numbers} \\
$\neg(\textit{exists} ~ b.(b|a ~ \& ~ b \ast b=a))$ \\
\hline
2 \\
3 \\
5 \\
6 \\
7 \\
8 \\
10 \\
\hline
\end{tabular}
}
\caption{Data tables for the concepts of prime and non-square numbers.}
\label{primeNonSquare}
\end{footnotesize}
\end{center}
\end{figure}

After this concept has been formed HR checks to see whether the data table is equivalent to, 
subsumed by, or subsumes another data table, or whether it is empty.  Assuming the 
concept of non-square numbers has been formed previously by HR, the data tables of both 
the concept of prime numbers and the concept of non-square numbers, shown in Figure
\ref{primeNonSquare}, are compared.

HR would immediately see that all of its prime numbers are also non-squares, and so conjectures that this
is true for all prime numbers.  That is, it will make the following implication conjecture:

\begin{center}
\begin{tabular}{ccc}
$\underbrace{2 ~ = ~ |\{b ~ : ~ b|a\}|}$ & $\Rightarrow$ & $\underbrace{\neg(\textit{exists} ~ b.(b|a ~ \& ~ b \ast b=a))}$ \\
{\small prime number} & & {\small non-square number} \\
\end{tabular}
\end{center}

\section{Automated theory formation for Event-B models with HR}
\label{sec:aft4eventB}

In this section we show how gluing invariant (\ref{GluingInvariant}) introduced in the example of \S\ref{EventB-Background} can be generated through the use of theory formation and, in particular, with the HR system. 

\subsection{Construction of conjectures in the domain of the Mondex system} \label{TFMondex}
HR's input consists of a set of core concepts that describe the domain.  With 
respect to Event-B models, these core concepts are represented by the 
state of the system, i.e. variables, and by the static information given in the context
of the model, i.e. constants and sets.  Furthermore, a concept is composed of a series of examples.  
Here, animation traces are used to provide HR with a list of examples for each of the concepts of an 
Event-B model.  As mentioned before, we use ProB \cite{ProB03} 
to animate the models.  For the purpose of the example, in Figure \ref{DomainConcepts}
we introduce some of the core concepts with their respective data tables -- which were generated
through the animation of the model with the ProB system.

\begin{figure}[h]
\begin{center}
\begin{footnotesize}
\subfigure{
\begin{tabular}{|c|}
\hline
\multicolumn{1}{|c|}{\textbf{state(A)}} \\
\hline
S0 \\ 
S1 \\ 
S2 \\ 
S3 \\ 
S4 \\ 
S5 \\ 
S6 \\ 
S7 \\ 
S8 \\ 
S9 \\ 
S10 \\ 
$\vdots$ \\
S58 \\ 
S59 \\ 
\hline
\end{tabular}
}
\subfigure{
\begin{tabular}{|c|}
\hline
\multicolumn{1}{|c|}{\textbf{status(A)}} \\
\hline
IDLEF \\ 
EPR \\ 
EPA \\ 
ABORTEPR \\ 
ABORTEPA \\ 
ENDF \\ 
IDLET \\ 
EPV \\ 
ABORTEPV \\ 
ENDT \\ 
\hline
\end{tabular}
}
\subfigure{
\begin{tabular}{|c|}
\hline
\multicolumn{1}{|c|}{\textbf{purseSet(A)}} \\
\hline
purse1 \\ 
purse2 \\ 
purse3 \\ 
purse4 \\ 
purse5 \\ 
\hline
\end{tabular}
}
\subfigure{
\begin{tabular}{|c|c|}
\hline
\multicolumn{2}{|c|}{\textbf{idleFP(A,B)}} \\
\hline
S5 & purse3 \\
S6 & purse3 \\
S6 & purse5 \\
S7 & purse5 \\
S8 & purse5 \\
S19 & purse4 \\
S25 & purse5 \\
S29 & purse1 \\
S30 & purse1 \\
S31 & purse1 \\
S38 & purse5 \\
S39 & purse5 \\
S40 & purse5 \\
S44 & purse5 \\
S45 & purse5 \\
S52 & purse5 \\
S53 & purse5 \\
S54 & purse5 \\
\hline
\end{tabular}
}
\subfigure{
\begin{tabular}{|c|c|c|}
\hline
\multicolumn{3}{|c|}{\textbf{statusF(A,B,C)}} \\
\hline
S5 & purse3 & IDLEF \\ 
S6 & purse3 & IDLEF \\ 
S6 & purse5 & IDLEF \\ 
S7 & purse3 & EPR \\ 
S7 & purse5 & IDLEF \\ 
S8 & purse3 & EPR \\ 
S8 & purse5 & IDLEF \\ 
$\vdots$ & $\vdots$ & $\vdots$ \\
S29 & purse1 & IDLEF \\ 
S29 & purse5 & ABORTEPR \\ 
S30 & purse1 & IDLEF \\ 
S30 & purse5 & ABORTEPR \\ 
S31 & purse1 & IDLEF \\ 
S31 & purse5 & ABORTEPR \\ 
$\vdots$ & $\vdots$ & $\vdots$ \\
S59 & purse1 & ABORTEPA \\ 
\hline
\end{tabular}
}
\caption{Core concepts supplied to HR}
\label{DomainConcepts}
\end{footnotesize}
\end{center}
\end{figure}

\begin{figure}[h]
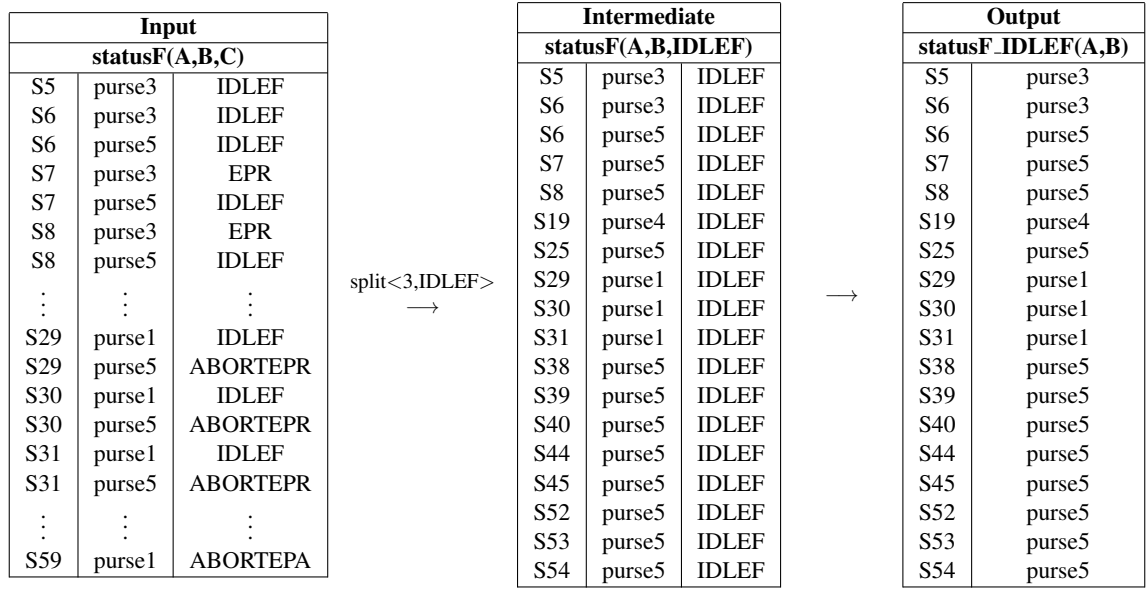

\begin{center}
\begin{footnotesize}
\subfigure{
\begin{tabular}{|c|c|c|}
\hline
\multicolumn{3}{|c|}{\textbf{Input}} \\
\hline
\multicolumn{3}{|c|}{\textbf{statusF(A,B,C)}} \\
\hline
S5 & purse3 & IDLEF \\ 
S6 & purse3 & IDLEF \\ 
S6 & purse5 & IDLEF \\ 
S7 & purse3 & EPR \\ 
S7 & purse5 & IDLEF \\ 
S8 & purse3 & EPR \\ 
S8 & purse5 & IDLEF \\ 
$\vdots$ & $\vdots$ & $\vdots$ \\
S29 & purse1 & IDLEF \\ 
S29 & purse5 & ABORTEPR \\ 
S30 & purse1 & IDLEF \\ 
S30 & purse5 & ABORTEPR \\ 
S31 & purse1 & IDLEF \\ 
S31 & purse5 & ABORTEPR \\ 
$\vdots$ & $\vdots$ & $\vdots$ \\
S59 & purse1 & ABORTEPA \\ 
\hline
\end{tabular}
}
\hspace{-3mm}
\subfigure{
\begin{scriptsize}
\begin{tabular}{c}
split$<$3,IDLEF$>$ \\
$\longrightarrow$ \\
\end{tabular}
\end{scriptsize}
}
\hspace{-3mm}
\subfigure{
\begin{tabular}{|c|c|c|}
\hline
\multicolumn{3}{|c|}{\textbf{Intermediate}} \\
\hline
\multicolumn{3}{|c|}{\textbf{statusF(A,B,IDLEF)}} \\
\hline
S5 & purse3 & IDLEF \\ 
S6 & purse3 & IDLEF \\ 
S6 & purse5 & IDLEF \\ 
S7 & purse5 & IDLEF \\ 
S8 & purse5 & IDLEF \\ 
S19 & purse4 & IDLEF \\ 
S25 & purse5 & IDLEF \\ 
S29 & purse1 & IDLEF \\ 
S30 & purse1 & IDLEF \\ 
S31 & purse1 & IDLEF \\ 
S38 & purse5 & IDLEF \\ 
S39 & purse5 & IDLEF \\ 
S40 & purse5 & IDLEF \\ 
S44 & purse5 & IDLEF \\ 
S45 & purse5 & IDLEF \\ 
S52 & purse5 & IDLEF \\ 
S53 & purse5 & IDLEF \\ 
S54 & purse5 & IDLEF \\ 
\hline
\end{tabular}
}
\hspace{2mm}
\subfigure{
\begin{scriptsize}
$\longrightarrow$
\end{scriptsize}
}
\hspace{2mm}
\subfigure{
\begin{tabular}{|c|c|}
\hline
\multicolumn{2}{|c|}{\textbf{Output}} \\
\hline
\multicolumn{2}{|c|}{\textbf{statusF\_IDLEF(A,B)}} \\
\hline
S5 & purse3 \\ 
S6 & purse3 \\ 
S6 & purse5 \\ 
S7 & purse5 \\ 
S8 & purse5 \\ 
S19 & purse4 \\ 
S25 & purse5 \\ 
S29 & purse1 \\ 
S30 & purse1 \\ 
S31 & purse1 \\ 
S38 & purse5 \\ 
S39 & purse5 \\ 
S40 & purse5 \\ 
S44 & purse5 \\ 
S45 & purse5 \\ 
S52 & purse5 \\ 
S53 & purse5 \\ 
S54 & purse5 \\ 
\hline
\end{tabular}
}
\caption{Split rule applied to obtain the concept of purses whose status in function \textit{statusF} is \textit{IDLEF}.}
\label{StatusIdleF}
\end{footnotesize}
\end{center}
\end{figure}

Then, HR applied all possible combinations of concepts and production rules in order to generate 
new concepts and form conjectures.  After the 433 step, HR formed the concept of the set of purses 
whose status in function \textit{statusF}  maps to \textit{IDLEF} by applying the \textit{split} production
rule.  The application of this step is illustrated in Figure \ref{StatusIdleF}.  An intermediate
output is generated with all tuples of concept \emph{statusF} whose third column matches the parameter
\emph{IDLEF}.  Since the third column is the same for all tuples of the intermediate concept, this
column is removed from the final output concept.
 
\begin{figure}[h!]
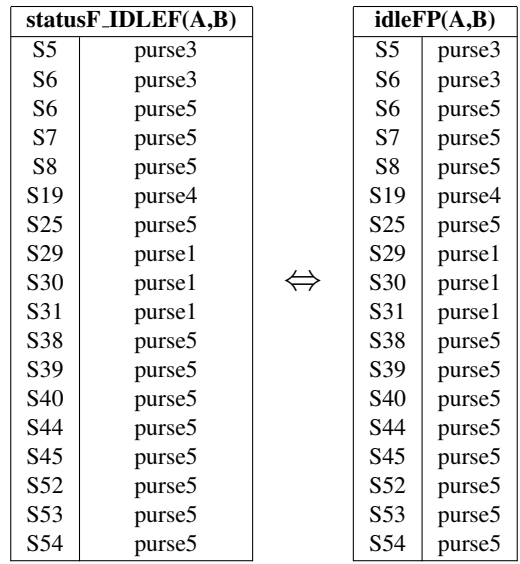

\begin{center}
\begin{footnotesize}
\subfigure{
\begin{tabular}{|c|c|}
\hline
\multicolumn{2}{|c|}{\textbf{statusF\_IDLEF(A,B)}} \\
\hline
S5 & purse3 \\ 
S6 & purse3 \\ 
S6 & purse5 \\ 
S7 & purse5 \\ 
S8 & purse5 \\ 
S19 & purse4 \\ 
S25 & purse5 \\ 
S29 & purse1 \\ 
S30 & purse1 \\ 
S31 & purse1 \\ 
S38 & purse5 \\ 
S39 & purse5 \\ 
S40 & purse5 \\ 
S44 & purse5 \\ 
S45 & purse5 \\ 
S52 & purse5 \\ 
S53 & purse5 \\ 
S54 & purse5 \\ 
\hline
\end{tabular}
}
\hspace{1mm}
\subfigure{
\begin{Large}$\Leftrightarrow$\end{Large}
}
\hspace{1mm}
\subfigure{
\begin{tabular}{|c|c|}
\hline
\multicolumn{2}{|c|}{\textbf{idleFP(A,B)}} \\
\hline
S5 & purse3 \\
S6 & purse3 \\
S6 & purse5 \\
S7 & purse5 \\
S8 & purse5 \\
S19 & purse4 \\
S25 & purse5 \\
S29 & purse1 \\
S30 & purse1 \\
S31 & purse1 \\
S38 & purse5 \\
S39 & purse5 \\
S40 & purse5 \\
S44 & purse5 \\
S45 & purse5 \\
S52 & purse5 \\
S53 & purse5 \\
S54 & purse5 \\
\hline
\end{tabular}
}
\caption{Formed equivalence conjecture.}
\label{Conjecture}
\end{footnotesize}
\end{center}
\end{figure}

Immediately after the generation of new concepts, HR looks for relationships with other existing concepts.  As shown in Figure \ref{Conjecture}, HR found that the new concept has the same list of examples as concept \emph{idleFP}, which gives rise to the following equivalence conjecture:

\begin{center} 
$\forall A,B.(\textit{state}(A) ~ \wedge ~ \textit{purseSet}(B) ~ \wedge ~ \textit{idleFP}(A,B) ~ \Leftrightarrow ~ \textit{status}(\textit{IDLEF}) ~ \wedge ~ \textit{statusF}(A,B,\textit{IDLEF}))$
\end{center}

which can be represented in Event-B as (\ref{GluingInvariant}).

\subsection{Challenges in applying HR}
For the domain of the Mondex system a total of 4545 conjectures were generated after 1000 formation steps.  As can be observed, this is a considerable set of conjectures to analyse.  In general, using HR for the discovery of invariants presented us with three main challenges:
\begin{enumerate}
\item The HR theory formation mechanism consists of an iterative application of production rules over existing and new concepts.  In order for HR to perform an exhaustive search, all possible combinations of production rules and concepts must be carried out.  However, there is not a fixed number of theory formation steps set up for this process, since this varies depending on the domain, i.e. some domains need more theory formation steps than others.  This represented a challenge for the use of HR in the discovery of invariants since it was possible that an invariant had not been formed only because not enough formation steps were run.
\item Some production rules are more effective in certain domains than others.  Selecting the appropriate production rules results in the construction of a more interesting theory.  For instance, if we are looking at a refinement step in an Event-B model that introduces a partition of sets we expect the new invariants to define properties over the new sets; therefore, production rules like the \emph{arithmetic} production rule will not be of much interest in the development of the theory associated to the refinement step.  Automatically selecting appropriate production rules requires knowledge about the domain; therefore, a technique was needed in order to perform this selection.
\item Finally, as highlighted in our example, HR produces a large number of conjectures -- in our experiments some where in the range of 3000 to 12000 conjectures per run -- from which only a very small set represented interesting invariants of the system.  Thus, our main challenge was to find a way of automatically selecting the conjectures that are interesting for the domain among the conjectures obtained from HR.
\end{enumerate}

In order to overcome these challenges, we have developed an approach that uses proof failure analysis to guide the search in HR.  In the next section, we introduce this approach and illustrate its application, based on our running example from the Mondex case study.

\section{Proof failure analysis and HR}
\label{sec:approach}
In order to use HR, a user must first configure the system for their 
application domain. This involves the user in selecting production rules
and conjecture making techniques, as well as deciding how many steps 
HR should be run. In the example introduced in \S\ref{TFMondex}, the application of the split
production rule with respect to the concept \emph{statusF}, for the value
\emph{IDLEF}, is an informed decision, based upon the user's knowledge of the model.
On its own, HR does not have the capability of applying this type of reasoning. 
Often particular combinations of these parameters turn out to be useful for
different domains. Finding the right combination is largely a process of trial
and error.

Here we have developed a heuristic approach with the aim of automating this
trial and error process.  Our heuristics exploit the strong interplay between
modelling and reasoning in Event-B.  In the context of the
discovery of invariants through theory formation, we use the feedback provided
by failed POs to make decisions about how to configure HR in order to guide 
the search for invariants.  Specifically, our approach consists of analysing the 
structure of failed POs so that we can automate:

\begin{enumerate*}
\item the prioritisation in the development of conjectures about specific concepts,
\item the selection of appropriate production rules that increase the possibilities
of producing the missing invariants and,
\item the filtering of the final set of conjectures to be analysed as possible candidate invariants.
\end{enumerate*} 

\subsection{Heuristics}
Our heuristics constrain the search for invariants
by focusing HR on concepts that occur within failed POs. We use two
classes of heuristics -- those used in configuring HR, i.e. 
\emph{Pre-Heuristics (PH)}, and those used in selecting conjectures from
HR's output, i.e. \emph{Selection Heuristics (SH)}. Below we explain each
class of heuristics in turn:

\subsubsection{HR configuration heuristics} \label{ConfHeuristics}
We use two overall heuristics when configuring HR for a given Event-B refinement:

\begin{description}
 \item[PH1.] \textit{Prioritise core and non-core concepts that occur within the failed POs as follows:}.
\begin{description}
 \item [Goal concepts:] concepts that appear within the goals of the failed POs.
 \item [Hypotheses concepts:] concepts that appear within the hypotheses of the failed POs.
 \item [Other concepts:] concepts that do not appear within the failed POs.
\end{description}
 \item[PH2.] \textit{Select production rules which will give rise to
     conjectures relating to the concepts occurring within the failed POs, i.e.}
\begin{description}
 \item[Split rule:] \textit{is selected if members of finite sets occur}.
 \item[Arithmetic rule:] \textit{is selected if there are occurrences of
     arithmetic operators, e.g. +,-,*,/}.
 \item[Numrelation rule:] \textit{is selected if there are occurrences of
     relational operators, e.g. $>$,$<$,$\leq$,$\geq$}.
\end{description}
 \textit{In addition, because of the set theoretic nature of Event-B, the
   compose, disjunct and negate production rules are always used in the search
   for invariants -- where compose relates to conjunction and intersection,
   disjunct relates to disjunction and union and negate relates to negation
   and set complement}.
\end{description}

Below we provide the rationale for these heuristics:
\begin{itemize}

\item As explained in \S\ref{HRBackground}, HR uses the agenda mechanism to
      organise the theory formation steps. The purpose of PH1 is to give higher 
      priority to core and non-core concepts that occur within the failed POs, 
      which means HR will generate related conjectures earlier within the 
      theory formation process by having the prioritised concepts in the top
      of the agenda.

      Furthermore, we have observed that in most cases, we are able to identify the missing 
      invariants by focusing in the first instance on the concepts that arise 
      within the goals of the failed POs.  As a result, such concepts are assigned the
      highest priority in the application of heuristic PH1. The concepts associated 
      to the hypotheses follow in order of interest, while the remaining concepts 
      are given the lesser priority. 

\item The missing invariants that are required in order to overcome proof
      failures will typically have strong syntactic similarities with the 
      failed POs. This is the intuition behind PH2, which selects production
      rules that focus HR's theory formation process on such syntactic similarities. 
\end{itemize}

As will be shown in \S\ref{sec:experiments}, the empirical evidence we have
gathered so far supports our rationale.

\subsubsection{Conjecture selection heuristics}
In order to prune the set of conjectures generated by HR, we use the
following five selection heuristics:
\begin{description}
 \item[SH1.] \textit{Select conjectures that focus purely on prioritised core and non-core concepts.}
 \item[SH2.] \textit{Select conjectures where the sets of variables occurring on the left- and right-hand sides are disjoint}.
 \item[SH3.] \textit{Select only the most general conjectures}.
 \item[SH4.] \textit{Select conjectures that discharge the failed POs}.
 \item[SH5.] \textit{Select conjectures that minimise the number of additional
     proof failures that are introduced}.
\end{description}
The rationale for these heuristics is as follows:
\begin{itemize}
\item SH1 initiates the pruning of uninteresting conjectures by selecting
      those that describe properties about the prioritised core and non-core
      concepts (as identified by PH1). Furthermore, the selected conjectures
      should focus purely on the prioritised concepts; this means that we are
      interested only in equivalence and implication conjectures of the forms: 
\begin{center}
$\alpha$ $\Leftrightarrow$ $\beta$ \\
$\alpha$ $\Rightarrow$ $\beta$ \\
$\beta$ $\Rightarrow$ $\alpha$ \\
\end{center}
      where $\alpha$ relates to a prioritised core or non-core concept and $\beta$ to any other concept. 
      All non-existence conjectures associated with the prioritised concepts are selected.
      Note that this selection criteria still gives rise to a large set of
      conjectures. However, as explained in the rationale of PH1 in \S\ref{ConfHeuristics},
      in most cases we have identified the missing invariants by focusing first on the concepts
      associated to the goals of the failed POs.  For the selection process the same reasoning is 
      followed and, therefore, heuristics SH1 to SH5 are focused first on conjectures associated 
      to the concepts of the goals identified by the application of PH1.  If no candidate invariants are found,
      or if old failures are still not addressed by the identified invariants, then the selection
      process starts again from SH1 to SH5 but focused on the conjectures associated with the 
      concepts of the hypotheses.

\item SH2 further prunes the set of conjectures by selecting only those that do not
      use the same variable(s) in both sides of the conjecture. The reason for
      this is that invariants in Event-B typically express relationships between different 
      variables of the model. 

\item SH3 is used to eliminate redundancies amongst the set of selected
      conjectures by removing those that are logically implied by more general 
      conjectures. 

\item SH4 is used to select candidate invariants which discharge the 
      given failed POs.

\item Potentially, overcoming one proof failure via the introduction of
      missing invariants may give rise to new proof fails. SH5 selects 
      conjectures that discharge the failed POs, whilst minimising the
      number of new failed POs that are introduced. This iterative 
      approach to discovering all the missing invariants is typical
      of Event-B developments, as described in Section 5 of \cite{Mondex}, 
      where invariant discovery is manual. Of course, if a
      development is incorrect, then this process will not terminate. We
      return to the issue of working with incorrect developments
      in \S\ref{sec:future}.  

\end{itemize}

Note that the selection conjectures must be applied in order from SH1 to SH5 
so as to optimise the selection procedure.

\subsection{Worked example}
We now illustrate the application of our heuristics by returning to the 
refinement step described in \S\ref{TFMondex}. Recall that the
gluing invariant (\ref{GluingInvariant}) was required in order for 
the correctness of the refinement to be proved.  When this invariant is 
missing from the model, an unprovable guard strengthening (\emph{GRD}) PO\footnote{
A GRD PO verifies that the guards of a refined event imply the guards of the
abstract event.}, as shown in Figure~\ref{Failure1}, is generated. The failed PO shows that the guard \emph{p1 $\in$ idleFP} of the abstract 
event is not implied by the guards of the concrete event.

\begin{figure}[!h]
\begin{center}
\begin{footnotesize}
\subfigure{
\begin{tabular}{|l|}
\hline
\textbf{Failed PO:} \\
\hline
p1 $\mapsto$ IDLEF $\in$ statusF \\
t $\in$ startFromM \\
p1 = from(t) \\
Fseqno(t) = currentSeqNo(p1) \\
$\vdash$ \\
p1 $\in$ idleFP \\
\hline
\end{tabular}
}
\caption{Failed GRD PO resulting from a missing gluing invariant}
\label{Failure1}
\end{footnotesize}
\end{center}
\end{figure}

We start the process of invariant discovery with the application of heuristic
PH1. We extract the list of core concepts that occur in the failed PO,
giving them higher priority within the theory formation process. The
extracted concepts are:

\begin{center}
\textit{idleFP, statusF, status, startFromM, from, FSeqno} and \textit{currentSeqNo}
\end{center}

Except for \emph{status}, all these concepts explicitly occur within 
the PO. Note that \emph{status} is added because the constant \emph{IDLEF} 
is a representative of the set \emph{status}.

Regarding non-core concepts, the hypothesis \emph{p1 $\mapsto$ IDLEF $\in$ statusF} in the failed PO suggests that function {\em statusF} 
maps an arbitrary purse to the status {\em IDLEF}.  This is an example of a non-core concept.
This concept is obtained through the application of the split production rule over the 
concept \emph{statusF} on the value \emph{IDLEF}.  No other non-core concepts are identified in the PO.

The next step is the selection of the production rules.  The following 
production rules are used in the invariant discovery process:

\begin{center}
\textit{compose, disjunct, negate} and \textit{split}
\end{center}

The compose, disjunct and negate production rules are always used in the
search, as defined by heuristic PH2. The split production rule is selected
because hypothesis \emph{p1 $\mapsto$ IDLEF $\in$ statusF} makes reference to
a member of the finite set \emph{status}: namely, the constant IDLEF.  Thus,
the split production rule is applied over the finite set \emph{status} and the
values to split are all the members of the set, i.e.: \emph{IDLEF, EPR, EPA,
ABORTEPR, ABORTEPA, ENDF, IDLET, EPV, ABORTEPV} and \emph{ENDT}.

After the application of the PH heuristics, the initial configuration of HR
is complete. By running HR for 1000 steps, 2134 conjectures were formed.
This should be compared with the 4545 conjectures that are generated 
if our PH heuristics are not used to configure HR. 

Now turning to the SH heuristics, SH1 selects conjectures that
relate to the prioritised concepts that appear within the goal of the failed PO.
In our example, this focuses on conjectures that involve the concept
\emph{idleFP}.  After applying SH1 we obtained:

\begin{center}
\textit{4 equivalences, 2 implications and 79 non-exists conjectures}
\end{center}

The application of SH2 removes conjectures whose left- and right-hand sides
are not disjoint with respect to the variable occurrences.  The application of SH2 yields the following results:

\begin{center}
\textit{1 equivalence, 2 implications and 79 non-exists conjectures}
\end{center}

Through the application of SH3, less general conjectures are removed. Applying this heuristic produces:

\begin{center}
\textit{1 equivalence, 2 implications and 46 non-exists conjectures}
\end{center}

SH4 selects only conjectures that discharge the failed PO, the results of this step are:

\begin{center}
\textit{1 equivalence, 0 implications and 0 non-exists conjectures}
\end{center}

Only one conjecture discharges the failed PO.  Furthermore, this conjecture 
does not introduce any additional failures; therefore, it represents an
invariant.  Within HR the invariant takes the form:

\begin{center}
$\forall A,B.(\textit{state}(A) ~ \wedge ~ \textit{purseSet}(B) ~ \wedge ~ \textit{idleFP}(A,B) ~ \Leftrightarrow ~ \textit{status}(\textit{IDLEF}) ~ \wedge ~ \textit{statusF}(A,B,\textit{IDLEF}))$
\end{center}

which translates into the missing gluing invariant (\ref{GluingInvariant}).
It should be noted that this conjecture was formed by HR after one theory
formation step. This shows that, in this example, our heuristics guided HR to discover
interesting conjectures early within the theory formation process.

\section{Experimental results}
\label{sec:experiments}

The experiments we carried out were divided into two stages. The first stage
involved the {\em development} of our heuristics, and was based upon four relatively
simple Event-B models, as described below:
\begin{enumerate}
 \item \emph{Traffic light system:} This model represents a traffic light
   circuit that controls the sequencing of lights.  It is composed of an
   abstract model and involves a single refinement. The abstract model controls the red and
   green lights, while the refinement introduces a third light to the
   sequence, i.e. an amber light.
 \item Two representations of a vending machine:
	\begin{itemize}
   	\item \emph{Set-like representation:} This model of a vending machine
          controls the stock of products through the use of states.  It is
          composed of an abstract and a concrete model.  The abstract model
          represents the states of products using state sets, while the refinement
          introduces a status function that maps products to their states. 
	\item \emph{Arithmetic-like representation:} This model of
          the vending machine uses natural numbers to represent the stock 
          and money held within the machine. While the abstract model 
          deals with a single product, the refinement introduces
          a second product to the vending machine. 
	\end{itemize}
 \item \emph{Refinements 1 and 2 of Abrial's cars on a bridge system \cite{EventbBook}:} Models
   a system that controls the flow of cars on a bridge that connects a
   mainland to an island.  At the abstract level, cars are modelled leaving and entering
   the island, the first refinement introduces the requirement that the bridge 
   only supports one way traffic, while the second refinement introduces traffic lights.
\end{enumerate}

We used the second stage of our experiments to {\em evaluate} the heuristics
developed during stage one. Here the experiments were performed on
more complex Event-B models:
\begin{enumerate}
 \item \emph{Refinement 3 of Abrial's cars on a bridge system \cite{EventbBook}:} The third
   refinement of this system models the introduction of sensors that detect
   the physical presence of cars. 
 \item \emph{The Mondex system \cite{Mondex}:} Models an electronic purse that allows the
   transfer of money between purses. This development is composed of one
   abstract model and nine refinement steps.  We targeted the
   third, fourth and eighth refinement steps.  The third refinement handles
   dual state sets in both sides of a transaction in order to handle
   information locally.  The fourth refinement introduces the use of messaging
   channels between purses and the eighth refinement introduces a status
   function that maps purses to their states instead of using state sets.
\end{enumerate}

In the work reported in \cite{Mondex}, it was highlighted that the manual
analysis of failed POs was used to guide the construction of gluing
invariants.  In particular, this was illustrated in the third step of the
refinement in which, through the analysis of failed POs, and after three
iterations of invariant strengthening, the set of invariants needed to prove
the refinement between levels three and four were added to the model.  As part of our
experiments we attempted the re-construction of the Mondex system in Event-B
based on the development presented in \cite{Mondex}.  In the following section
we present the results obtained by the application of our approach to the
refinement between levels three and four of the Mondex system, and we show that these results are
similar to the ones obtained through the interactive development \cite{Mondex}.

\subsection{The Mondex system} 
\label{Ref3}
In level three of the Mondex system a transaction is permitted to be in one of four
states: \emph{idle, pending, recover or ended}, while the refinement in level
four introduces dual states to a transaction so that each side has their own
local protocol state.  In order to evaluate our approach, we introduced the
model in level 4 with only basic typing invariants.  The absence of the
invariants produces the failed POs shown in Figure \ref{MondexPOs1}.

\begin{figure}[h]
\begin{center}
\begin{footnotesize}
\subfigure{
\begin{tabular}{|l|}
\hline
\textbf{PO1:} \\
\hline
p1 $\in$ purse \\
t $\in$ epr \\
t $\in$ epv $\cup$ abortepv \\
p1 = from(t) \\
a $\in$ $\mathbb{N}$ \\
a = am(t) \\
a $\leq$ bal(p1) \\
$\vdash$ \\
t $\in$ idle \\
\hline
\end{tabular}
}
\subfigure{
\begin{tabular}{|l|}
\hline
\textbf{PO2:} \\
\hline
p1 $\in$ purse \\
p2 $\in$ purse \\
t $\in$ epv \\
t $\in$ epa $\cup$ abortepa \\
a $\in$ $\mathbb{N}$ \\
a = am(t) \\
p1 = from(t) \\
p2 = to(t) \\
$\vdash$ \\
t $\in$ pending \\
\hline
\end{tabular}
}
\subfigure{
\begin{tabular}{|l|}
\hline
\textbf{PO3:} \\
\hline
t $\in$ epv \\
t $\in$ abortepa \\
$\vdash$ \\
t $\in$ pending \\
\hline
\multicolumn{1}{c}{} \\
\hline
\textbf{PO4:} \\
\hline
t $\in$ epa \\
t $\in$ abortepv \\
$\vdash$ \\
t $\in$ pending \\
\hline
\end{tabular}
}
\subfigure{
\begin{tabular}{|l|}
\hline
\textbf{PO5:} \\
\hline
p1 $\in$ purse \\
t $\in$ abortepa \\
t $\in$ abortepv \\
a $\in$ $\mathbb{N}$ \\
a = am(t) \\
p1 = from(t) \\
$\vdash$ \\
t $\in$ recover \\
\hline
\end{tabular}
}

\caption{First set of failed POs.}
\label{MondexPOs1}
\end{footnotesize}
\end{center}
\end{figure}

We start the invariant discovery process with the application of heuristic PH1.  The set of core concepts selected from the failed POs are:

\begin{center}
\emph{idle, pending, recover, purse, epr, epv, abortepv, from, am, bal, epa, abortepa} and \emph{to}
\end{center}

Moreover, from the analysis of the predicates in the failed POs, we identify the following non-core concepts:

\begin{center}
\emph{epv $\cup$ abortepv} and \emph{epa $\cup$ abortepa}
\end{center}

These concepts are identified from hypotheses \emph{t $\in$ epv $\cup$ abortepv} 
and \emph{t $\in$ epa $\cup$ abortepa} within PO1 and PO2, respectively.  Note that 
\emph{t} does not represent a concept in the domain, it represents an arbitrary 
transaction passed as a parameter to the event associated with the failed POs.  For this 
reason, only the right hand sides of the membership relations are selected as
interesting non-core concepts. 

The process continues with the selection of the productions rules.  Based on
the failed POs shown in Figure \ref{MondexPOs1}, the following production
rules are selected for the search:

\begin{center}
\emph{compose, disjunct, negate} and \emph{numrelation}
\end{center}

The compose, disjunct and negate production rules are always used in the
search as stated in heuristic PH2.  The numrelation production rule is
selected because hypothesis $a \leq bal(p1)$ within PO1 expresses a property based
on the relational operator $\leq$.  After the pre-heuristics have been applied HR is 
run for 1000 steps, resulting in 7296 conjectures.  

The selection heuristics are now applied over this set of conjectures.
Heuristic SH1 suggests looking at the prioritised concepts associated to the
goals of the failed POs.  From the goals of the POs shown in Figure
\ref{MondexPOs1}, we identified the concepts \emph{idle, pending} and
\emph{recover}.  Thus, we look for the conjectures associated to each of these
concepts.  The results from the application of this heuristic are shown in
Table \ref{MonL4}.  This table also shows the results of applying heuristics
SH2, SH3 and SH4 over each of the selected concepts.

\begin{table}[h!]
\begin{footnotesize}
\begin{center}
\begin{tabular}{|c|l|c|c|c|}
\hline
\multicolumn{1}{|c|}{\textbf{Heuristic}} & \multicolumn{1}{|c|}{\textbf{Concept}} & \textbf{Equivalences} & \textbf{Implications} & \textbf{Non-exists} \\
\hline
\multirow{3}{*}{SH1} & idle & 7 & 27 & 24 \\
& pending & 6 & 27 & 35 \\
& recover & 9 & 51 & 41 \\
\hline
\multirow{3}{*}{SH2} & idle & 0 & 27 & 24 \\
& pending & 0 & 27 & 35 \\
& recover & 2 & 51 & 41 \\
\hline
\multirow{3}{*}{SH3} & idle & 0 & 6 & 17 \\
& pending & 0 & 8 & 26 \\
& recover & 2 & 3 & 30 \\
\hline
\multirow{3}{*}{SH4} & idle & 0 & 2 & 0 \\
& pending & 0 & 2 & 0 \\
& recover & 1 & 0 & 0 \\
\hline
\end{tabular}
\end{center}
\caption{Results of the application of selection heuristics SH1, SH2, SH3 and SH4.}
\label{MonL4}
\end{footnotesize}
\end{table}

As can be observed, after applying the four initial selection heuristics we
have narrowed the set of selected conjectures to a total of five conjectures:
two implications involving the concept \emph{idle}, two implications for concept
\emph{pending} and one equivalence about the concept \emph{recover}. 

The final step in the discovery process is the selection of the conjectures
that produce the smaller number of new failed POs.  The two implications
associated with concept \emph{idle} discharge PO1 and produce one extra failed
PO.  We believe that in this kind of situation it is the user who has to decide
which one is the most appropriate conjecture according to his/her knowledge about
the model.  Thus, we present both conjectures as candidate invariants and
leave the decision of which one to select to the user.  Regarding the two implications 
associated with concept \emph{pending}, one of them discharges PO2 and PO3 
and produces two new failed POs, while the other one discharges PO4 but produces 
three new failed POs.  As there are no other conjectures that help to overcome the 
failures produced by PO2, PO3 and PO4, both conjectures are suggested as candidate 
invariants.  Finally, the equivalence conjecture associated with concept \emph{recover} 
discharges PO5 and it does not produce any extra failures, so this conjecture is also 
suggested as a candidate invariant.  The set of invariants represented by the conjectures obtained from
HR in this first iteration of our approach is shown in Figure \ref{InvM4I1} \footnote{Note that 
we have given the equivalent set theoretic representation of these conjectures
instead of using the universally quantified format provided by HR. This is because some experiments, 
for instance the development of the Mondex system carried out in \cite{Mondex}, have
shown that the automatic provers do better with quantifier-free predicates.}.

After the new set of invariants is introduced to the model, six new failed POs
are generated.  We then start the analysis again by applying our approach
based on the new set of failed POs.  This new iteration results in the 
discovery of five new invariants.  Again, when these invariants are added to
the model, one new failed PO is generated.  We discovered one new invariant after
a third iteration of our approach.  No further failed POs are generated when
this invariant is added to the model.  Figures \ref{InvM4I2} and Figure
\ref{InvM4I3} shows the invariants obtained after the second and the third
iteration, respectively.   

\begin{figure}[h!]
\begin{footnotesize}
\begin{center}
\subfigure[]{
\begin{tabular}{l}
epr $\subseteq$ idle \\
(idleF $\cup$ epr) $\subseteq$ idle \\
epv $\cap$ (epa $\cup$ abortepa) $\subseteq$ pending \\ 
epa $\cap$ (epv $\cup$ abortepv) $\subseteq$ pending \\ 
abortepa $\cap$ abortepv = recover \\ 
\end{tabular}
\label{InvM4I1}
}
\subfigure[]{
\begin{tabular}{l}
idleF $\subseteq$ idle \\
idleT $\cap$ (epa $\cup$ abortepa) $=$ $\emptyset$ \\ 
idleT $\cap$ (epv $\cup$ abortepv) $=$ $\emptyset$ \\ 
epv $\cap$ abortepv $=$ $\emptyset$ \\ 
epa $\cap$ abortepa $=$ $\emptyset$ \\ 
\end{tabular}
\label{InvM4I2}
}
\subfigure[]{
\begin{tabular}{l}
epr $\cap$ idleF $=$ $\emptyset$ \\ 
\end{tabular}
\label{InvM4I3}
}
\end{center}
\vspace{-5mm}
\caption{Invariants obtained through three iterations.}
\label{InvR4}
\end{footnotesize}
\end{figure}

The invariants shown in Figure \ref{InvR4} are a subset of the invariants
suggested in \cite{Mondex} for this step of the refinement.  In total we
obtained 11 invariants from the 17 used in \cite{Mondex}.  However, it is
important to note that we have addressed all the failures produced when
proving consistency between the refinement levels.  Our hypothesis, is that
the extra invariants used in \cite{Mondex} represent new requirements of the
system, which are out of the scope of our technique since we only target 
invariants needed to prove the refinement steps.

\subsection{Summary of results}

Table \ref{Summary} summarises the results of the application of our approach
in each of the Event-B models used during the development and the evaluation
stages.  Notice that all the experiments were performed over models with only
basic typing invariants.   This means that neither gluing nor system
invariants were present in the models when using our technique.
The table shows for each refinement step, the number of failed POs that arose, 
as well as the number of gluing and system invariants discovered through our 
approach.  We also record the number of iterations involved in the invariant
discovery process.

\begin{table}[h!]
\begin{center}
\begin{footnotesize}
\begin{tabular}{c|c|c|c|c|c|c|c|}
\cline{2-8}
& \multirow{2}{*}{\textbf{Event-B model}} & \multirow{2}{*}{\textbf{Step}} & \multirow{2}{*}{\textbf{Failed POs}} & \multicolumn{4}{|c|}{\textbf{Invariants}} \\
\cline{5-8}
& & & & \multicolumn{4}{|c|}{\textbf{Automatically discovered}} \\
\cline{5-8}
& & & & \textbf{Glue} & \textbf{System} & \textbf{Total} & \textbf{Iteration} \\
\cline{2-8}
\multirow{5}{*}{Development set} & Traffic light & Level 1-2 & 2 & 2 & 0 & \textbf{2} & 1 \\
\cline{2-8}
& Vending machine (Arith) & Level 1-2 & 6 & 3 & 0 & \textbf{3} & 1 \\
\cline{2-8}
& Vending machine (sets) & Level 1-2 & 6 & 3 & 0 & \textbf{3} & 1 \\
\cline{2-8}
& \multirow{3}{*}{Cars on a bridge} & Level 1-2 & 2 & 1 & 0 & \textbf{1} & 1 \\
& & Level 2-3 & 6 & 0 & 5 & \textbf{5} & 1 \\
\hdashline
\multirow{8}{*}{Evaluation set} & & Level 3-4 & 7 & 0 & 5 & \textbf{5} & 1 \\
\cline{2-8}
& \multirow{7}{*}{Mondex} & \multirow{3}{*}{Level 3-4} & 5 & 4 & 0 & \textbf{4} & 1  \\
& & & 6 & 1 & 4 & \textbf{5} & 2 \\
& & & 1 & 0 & 1 & \textbf{1} & 3 \\
\cline{3-8}
& & \multirow{3}{*}{Level 4-5} & 3 & 0 & 3 & \textbf{3} & 1 \\
& & & 5 & 0 & 4 & \textbf{4} & 2 \\
& & & 4 & 0 & 2 & \textbf{2} & 3 \\
\cline{3-8}
& & Level 8-9 & 14 & 10 & 0 & \textbf{10} & 1 \\
\cline{2-8}
\end{tabular}
\caption{Automatically discovered invariants.}
\label{Summary}
\end{footnotesize}
\end{center}
\end{table}

In Table \ref{Compare} we compare our results with the actual
invariants given in the literature for the models of the cars on a bridge \cite{EventbBook}
and the Mondex system \cite{Mondex}; the other developments are not compared 
because they are of our own authorship (note that the invariants of the refinement from 
levels four to five of the Mondex system are not given in the literature). All automatically 
discovered invariants are subsets of the invariants given in the literature; however, it is 
important to highlight that the automatically discovered invariants were sufficient to prove 
all the refinement steps in our experimental models.

As it can be observed from Table \ref{Compare}, the automatic discovery of gluing invariants 
through the use of theory formation and the HR system has provided promising results. In 
most cases, the set of gluing invariants discovered through our technique was almost identical 
to the set of gluing invariants provided in the literature.  Regarding system invariants, it can 
be observed that the last refinement of the cars on a bridge system shows a big gap between 
the invariants given in the literature and those found automatically with our approach. As mentioned
previously, we believe that this difference can be explained by the introduction of new requirements, 
resulting in the need for extra properties in the model.

\begin{table}[h!]
\begin{center}
\begin{footnotesize}
\begin{tabular}{|c|c|c|c|c|c|c|c|}
\hline
\multirow{2}{*}{\textbf{Event-B model}} & \multirow{2}{*}{\textbf{Step}} & \multicolumn{3}{|c|}{\textbf{Given in Literature}} & \multicolumn{3}{|c|}{\textbf{Automatically discovered}} \\
\cline{3-8}
& & \textbf{Glue} & \textbf{System} & \textbf{Total} & \textbf{Glue} & \textbf{System} & \textbf{Total} \\
\hline
\multirow{3}{*}{Cars on a bridge} & Level 1-2 & 1 & 1 & \textbf{2} & 1 & 0 & \textbf{1} \\
\cline{2-8}
& Level 2-3 & 0 & 6 & \textbf{6} & 0 & 5 & \textbf{5} \\
\cline{2-8}
& Level 3-4 & 0 & 23 & \textbf{23} & 0 & 5 & \textbf{5} \\
\hline
\multirow{3}{*}{Mondex} & Level 3-4 & 8 & 9 & \textbf{17} & 5 & 5 & \textbf{10} \\
\cline{2-8}
& Level 4-5 & - & - & - & 0 & 9 & \textbf{9} \\
\cline{2-8}
& Level 8-9 & 10 & 0 & \textbf{10} & 10 & 0 & \textbf{10} \\
\hline
\end{tabular}
\label{Obtained}
\caption{Comparison between hand-crafted and automatically discovered invariants.}
\label{Compare}
\end{footnotesize}
\end{center}
\end{table}

Figure \ref{Invariants}, shows all the invariants that were discovered through
the application of our approach.  The invariants for refinement three of the Mondex
system are omitted since they are shown in \S\ref{Ref3}.

 \begin{figure}[h!]
 \begin{center}
 \begin{footnotesize}
 \subfigure[Vending machine (sets) invariants]{
 \begin{tabular}{l}
 available = productStatus$^{-1}$[$\{$AVAILABLE$\}$] \\
 limited = productStatus$^{-1}$[$\{$LIMITED$\}$] \\
 soldOut = productStatus$^{-1}$[$\{$SOLDOUT$\}$] \\
 \end{tabular}
 }
 \subfigure[Vending machine (arith) invariants]{
 \begin{tabular}{l}
 stock = stockMilk + stockPlain \\
 sold = soldMilk + soldPlain \\
 givenCoin = EMPTY\_COIN $\Leftrightarrow$ coin = NO\_COIN \\
 \end{tabular}
 }
 \subfigure[Cars on a bridge invariants]{
 \begin{tabular}{l}
 n = a + b + c \\
 ml\_tl = green $\Rightarrow$ c = 0 \\
 il\_tl = green $\Rightarrow$ a = 0 \\
 ml\_tl = red $\Rightarrow$ ml\_pass = 1 \\
 il\_tl = red $\Rightarrow$ il\_pass = 1 \\
 il\_tl = green $\Rightarrow$ ml\_tl = red \\
 ml\_out\_10 = TRUE $\Rightarrow$ ml\_tl = green \\
 il\_out\_10 = TRUE $\Rightarrow$ il\_tl = green \\
 IL\_IN\_SR = on $\Rightarrow$ A $>$ 0 \\
 IL\_OUT\_SR = on $\Rightarrow$ B $>$ 0 \\
 ML\_IN\_SR = on $\Rightarrow$ C $>$ 0 \\
 \end{tabular}
 }
 \subfigure[Mondex invariants (ref 4)]{
 \begin{tabular}{l}
 epr $\cap$ reqM $\subseteq$ epv $\cup$ abortepv \\
 epv $\cap$ valM $\subseteq$ epa $\cup$ abortepa \\
 endT = endF $\cup$ ackM \\
 reqM $\cap$ idleF $\subseteq$ epv $\cup$ abortepv \\
 valM $\cap$ idleT = $\emptyset$ \\
 epr $\cap$ valM = $\emptyset$ \\
 epr $\cap$ abortepa = $\emptyset$ \\
 valM $\cap$ idleF = $\emptyset$ \\
 abortepa $\cap$ idleF = $\emptyset$ \\
 \end{tabular}
 }
 \subfigure[Mondex invariants (ref 8)]{
 \begin{tabular}{l}
 idleFP = statusF$^{-1}$[$\{$IDLEF$\}$] \\
 eprP = statusF$^{-1}$[$\{$EPR$\}$] \\
 epaP = statusF$^{-1}$[$\{$EPA$\}$] \\
 aborteprP = statusF$^{-1}$[$\{$ABORTEPR$\}$] \\
 abortepaP = statusF$^{-1}$[$\{$ABORTEPA$\}$] \\
 endFP = statusF$^{-1}$[$\{$ENDF$\}$] \\
 idleTP = statusF$^{-1}$[$\{$IDLET$\}$] \\
 epvP = statusF$^{-1}$[$\{$EPV$\}$] \\
 abortepvP = statusF$^{-1}$[$\{$ABORTEPV$\}$] \\
 endTP = statusF$^{-1}$[$\{$ENDT$\}$] \\
 \end{tabular}
 }
 \subfigure[Traffic light invariants]{
 \begin{tabular}{l}
 r\_light = TRUE $\vee$ amber\_light = TRUE $\Leftrightarrow$ red\_light = TRUE \\
 g\_light = TRUE $\Leftrightarrow$ green\_light = TRUE \\
 \end{tabular}
 }
 \caption{Automatically discovered invariants}
 \label{Invariants}
 \end{footnotesize}
 \end{center}
 \end{figure}

\section{Related and future work}
\label{sec:future}

As far as we are aware, automated theory formation techniques have
not been investigated within the context of refinement style 
formal modelling. The closest work we know of is Daikon \cite{ErnstPGMPTX2007}, a 
system which uses templates to detect likely program 
invariants by analysing program execution traces. 
The quality of the invariants generated by both approaches depends in part upon
the quality of the input data -- ProB animation traces in our work and 
program test suites for Daikon. Like HR, Daikon is configurable.
However, while HR is a general purpose theory formation tool, Daikon 
has been designed with program analysis in mind. It should also be stressed 
that Daikon is a system, whereas the work presented here is
an initial investigation into developing an invariant generation 
tool for refinement based formal methods. 

Within automated theory formation there are a
number of alternative tools to HR that could be explored. For instance,
IsaCoSy \cite{JoDiBu10b}, IsaScheme \cite{Montano:MICAI:2010}, 
the CORE system \cite{Maclean2009:WING} and MathsAid \cite{mcca-bund-aute-at65}.
Underlying the first three of these systems is a notion of {\em term
  synthesis}, i.e. the automatic generation of 
candidate conjectures based upon application of domain knowledge. 
IsaCoSy and IsaScheme  
support the discovery of theorems within the context of mathematical induction, 
while MathsAid provides broader support for
the development of mathematical theories. The CORE system has a strong 
software verification focus, supporting the automatic generation of frame and 
loop invariants for use in reasoning about pointer programs. What distinguishes these approaches from 
HR is that they do not rely upon animation/execution traces, instead they follow
a generate-and-test approach, where the ``test'' component involves theorem
proving. Coupled with its configurability, the trace analysis capability led 
us to use HR for our investigations. 

As noted above, animation is  key to our approach, where the quality of 
the invariants produced by HR strongly depends on the quality of the 
animation traces. The ProB animator provided good animation traces for 
most of our experiments; however, we found two areas where further
improvements are required: 
\begin{enumerate}
 \item We believe that increasing the randomness in the production of the traces would
       improve our results.
 \item ProB preferences only allows for the creation of sets with a few elements, as
   well as very limited integer ranges.  This restricted some of the traces we 
   were able to generate, and thus impacted negatively on the invariants
   that could be discovered. Specifically, this limitation arose during 
   our analysis of the Mondex case study.
\end{enumerate}

The process of finding a ``correct'' refinement will typically involve
exploring many ``incorrect'' refinements. While the work reported here
focuses on supporting the verification of correct refinements, we are
currently investigating how counter-examples generated by ProB could
be combined with HR in order to provide useful feedback to a developer
when faced with an incorrect refinement. 

Longer-term, we are looking to use theory formation within our REMO \cite{IrelandGrovLlanoButler10}
formal modelling planning system. That is, when faced with a refinement 
failure, we aim to use theory formation, automatically tailored by 
refinement plans \cite{afm10}, to suggest modelling alternatives. 
Of course, such ``modelling alternatives''
are only suggestions, ultimately users must select which is most 
appropriate to their needs. 

Currently, the animation traces obtained from the ProB animator are automatically 
converted into HR's input, i.e. a domain file with the list of examples for each of 
the concepts (variables, constants and sets) is created.  However, the automation of 
the heuristics is still under development.  Automating the configuration 
heuristics involves the prioritisation of concepts in the domain file 
and the creation of a macro file.  The macro file records the search strategy that 
will be applied by HR (usually supplied by the user).  Here is where the production rules 
selected by our proof failure analysis are specified.  The automation of the selection 
heuristics requires integration with a theorem prover and the Rodin toolset.  HR uses 
the Otter theorem prover \cite{Otter} to prove the conjectures.  We will exploit the use 
of the Otter theorem prover in HR for the selection of the most general conjectures 
(heuristic SH3), while the Rodin toolset will be used to obtain the status of the POs 
after the candidate invariants have been introduced into the model (heuristics SH4 and SH5).
As future work, we aim to automate this process and, as mentioned before, integrate it with our REMO tool.

\section{Conclusions}
\label{sec:conclusions}
We have described an investigation into how the HR theory formation
tool can be used to automatically discover the kinds of invariants 
that developers typically have to supply in order to verify Event-B refinements. 
The key contribution of our work is the development of a
set of heuristics. Using proof-failure analysis to prune the wealth of 
conjectures HR discovers, these heuristics have proven highly effective 
at identifying missing invariants. While more experimentation is required, we believe that our heuristics 
provide a firm foundation upon which to further explore techniques that support 
formal refinement -- techniques that suggest design alternatives, whilst removing 
the burden of proof failure analysis from developers.

\vspace*{0.1in}

{\bf Acknowledgements:}
Our thanks go to Alan Bundy, Gudmund Grov and Julian Gutierrez
for their feedback and encouragement with this work.   Also, we want to thank Jens Bendisposto
and the ProB development team for their assistance, and Simon Colton and John Charnley
for their help in using the HR system.

\nocite{*}
\bibliographystyle{eptcs}
\bibliography{ref}

\end{document}